\documentclass{article}

\usepackage{arxiv}

\usepackage[utf8]{inputenc} 
\usepackage[T1]{fontenc}    
\usepackage{hyperref}       
\usepackage{url}            
\usepackage{booktabs}       
\usepackage{amsfonts}       
\usepackage{nicefrac}       
\usepackage{microtype}      
\usepackage{lipsum}		
\usepackage{graphicx}
\usepackage{natbib}
\usepackage{doi}
\usepackage{multirow}
\usepackage{multicol}
\usepackage{amsmath}

\title{Real-time EEG-based Emotion Recognition Model using Principal Component Analysis and Tree-based Models for Neurohumanities}

\author{\href{https://orcid.org/0009-0000-8621-578X}{\includegraphics[scale=0.08]{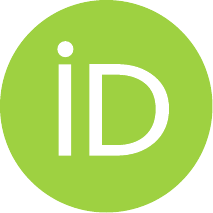}\hspace{1mm}Miguel A. Blanco-R\'{i}os\thanks{Equal contribution.}}\\
Department of Mechatronics\\
Tecnol\'{o}gico de Monterrey\\
Monterrey, Mexico \\
\texttt{A01284397@exatec.tec.mx} \\
\And
\href{https://orcid.org/0000-0001-7807-2539}{\includegraphics[scale=0.08]{orcid.pdf}\hspace{1mm}Milton O. Candela-Leal$^\ast$} \\
Department of Mechatronics\\
Tecnol\'{o}gico de Monterrey\\
Monterrey, Mexico \\
\texttt{milton.candela@tec.mx} \\
\And
\hspace{1mm}Cecilia Orozco-Romo \\
Department of Mechatronics\\
Tecnol\'{o}gico de Monterrey\\
Monterrey, Mexico \\
\texttt{A00827013@exatec.tec.mx} \\
\And
\hspace{1mm}Paulina Remis-Serna \\
Department of Mechatronics\\
Tecnol\'{o}gico de Monterrey\\
Monterrey, Mexico \\
\texttt{A00829183@exatec.tec.mx} \\
\And
Carol S. V\'{e}lez-Saboy\'{a} \\
Department of Humanistic Studies\\
Tecnol\'{o}gico de Monterrey\\
Monterrey, Mexico \\
\texttt{carol.velez@tec.mx} \\
\And
\href{https://orcid.org/0000-0001-5536-1426}{\includegraphics[scale=0.08]{orcid.pdf}\hspace{1mm}Jorge De-J. Lozoya-Santos} \\
Department of Mechatronics\\
Tecnol\'{o}gico de Monterrey\\
Monterrey, Mexico \\
\texttt{jorge.lozoya@tec.mx} \\
\And
\href{https://orcid.org/0000-0001-6359-2427}{\includegraphics[scale=0.08]{orcid.pdf}\hspace{1mm}Manuel Cebral-Loureda} \\
Department of Humanistic Studies\\
Tecnol\'{o}gico de Monterrey\\
Monterrey, Mexico \\
\texttt{manuel.cebral@tec.mx} \\
\And
\href{https://orcid.org/0000-0003-0306-2971}{\includegraphics[scale=0.08]{orcid.pdf}\hspace{1mm}Mauricio A. Ram\'{i}rez-Moreno\thanks{Corresponding author.}} \\
Department of Mechatronics\\
Tecnol\'{o}gico de Monterrey\\
Monterrey, Mexico \\
\texttt{mauricio.ramirezm@tec.mx} \\
}

\date{}

\renewcommand{\headeright}{}
\renewcommand{\undertitle}{}
\renewcommand{\shorttitle}{Real-time EEG-based Emotion Recognition for Neurohumanities}

\hypersetup{
pdftitle={Real-time EEG-based Emotion Recognition for Neurohumanities},
pdfsubject={eess.SP, cs.LG, q-bio.NC},
pdfauthor={Miguel A. Blanco-Rios, Milton O. Candela-Leal, Cecilia Orozco-Romo, Paulina Remis-Serna, Carol S. Velez-Saboya, Jorge De-J. Lozoya-Santos, Manuel Cebral-Loureda, Mauricio A. Ramirez-Moreno},
pdfkeywords={EEG, emotion recognition, real-time, PCA, random forest, humanities, neurohumanities, Descartes},
}

\begin{document}
\maketitle

\begin{abstract}
	Within the field of Humanities, there is a recognized need for educational innovation, as there are currently no reported tools available that enable individuals to interact with their environment to create an enhanced learning experience in the humanities (e.g., immersive spaces). This project proposes a solution to address this gap by integrating technology and promoting the development of teaching methodologies in the humanities, specifically by incorporating emotional monitoring during the learning process of humanistic context inside an immersive space. In order to achieve this goal, a real-time emotion detection EEG-based system was developed to interpret and classify specific emotions. These emotions aligned with the early proposal by Descartes (Passions), including admiration, love, hate, desire, joy, and sadness. This system aims to integrate emotional data into the Neurohumanities Lab interactive platform, creating a comprehensive and immersive learning environment. This work developed a ML, real-time emotion detection model that provided Valence, Arousal, and Dominance (VAD) estimations every 5 seconds. Using PCA, PSD, RF, and Extra-Trees, the best 8 channels and their respective best band powers were extracted; furthermore, multiple models were evaluated using shift-based data division and cross-validations. After assessing their performance, Extra-Trees achieved a general accuracy of 96\%, higher than the reported in the literature (88\% accuracy). The proposed model provided real-time predictions of VAD variables and was adapted to classify Descartes' six main passions. However, with the VAD values obtained, more than 15 emotions can be classified (reported in the VAD emotion mapping) and extend the range of this application.
\end{abstract}


\section{Introduction}
    An emotion is a psycho-physiological process triggered by the conscious or unconscious perception of an object or situation. This process is associated with a broad range of feelings, thoughts, and behaviors~\citep{Jarimowicz2006}. The study of emotion generation is pivotal, as it underpins the human experience, influencing cognition, perception, and daily activities, including learning, communication, and rational decision-making~\citep{Koelstra, mikhail}.

    Among the early descriptions of emotion, one is provided by Descartes, which he described as \emph{passions} in his work 'The Six Passions of Descartes.' For Descartes, the passions/emotions were experiences of the body on the soul, who, applying his famous method to moral philosophy, represented the problem of the passions of the soul in terms of its simplest integral components, distinguishing six different fundamental passions: admiration, love, hate, desire, joy, and sadness~\citep{descartes}.
   
    According to Descartes, admiration is understood as a sudden surprise of the soul that makes it consider (carefully) objects perceived as rare and extraordinary. This passion is directly related to the search for knowledge and philosophical reflection; love can be interpreted as the origin of the desire for union with someone or something that seems to be convenient; hate leads or drives to the rejection of something or someone; desire leads to an urge of possessing something that is out of reach; joy manifests when someone obtains that which they desire, while under a pleasant situation; lastly, sadness is experienced when losing something desired or during the experience of a painful situation~\citep{descartes}.

    The study of emotion has evolved over the years. While early definitions of emotions characterized them as bodily phenomena, modern approaches also acknowledge a cognitive component~\citep{Dixon2012}. The concept of emotion has been delved into both in the Humanistic and Scientific domains.

    \citet{humanities} define the humanities as encompassing all facets of human society and culture, including language, literature, philosophy, law, politics, religion, art, history, and social psychology. They underscore the importance of establishing closer collaborations between specific scientific domains, such as Neuroscience and the humanities. They posit that these collaborations will be mutually enriching and usher in a new era of profound and influential academic endeavors~\citep{humanities}. Hartley and Poeppel contend that advancements in theoretical, computational, neuroimaging, and experimental psychology have enabled linguistics, music, and emotion to emerge as central pillars of contemporary cognitive neuroscience~\citep{humanities2}.

   In education, the advancement of Humanities teaching methodologies has been notably slower compared to other fields, such as Science and Engineering~\citep{Cebral}. This lag underscores the necessity of integrating technology into the Humanities, given its potential to foster human flourishing and enhance emotional education. Recent studies suggest that immersive and interactive teaching environments can significantly improve learning experiences and outcomes~\citep{Chen}. Moreover, the infusion of neuroscience principles into pedagogical strategies is a burgeoning area of interest, with preliminary results indicating promising potential for improved educational experiences~\citep{Wilcox2021}.
   
   Developing research-based teaching approaches that combine Neuroscience and Education can help implement immersive and interactive systems for the teaching of Humanities. With this in mind, the \emph{Neurohumanities Lab} project emerges, which intends to implement an immersive and interactive platform for the education of humanities that allows users to interact with the environment (the classroom) and fosters impactful, logical and intuitive learning~\citep{Cebral2}. The idea behind the Neurohumanities Lab is to integrate a  system that can detect movements, actions, emotions, and physiological and mental states through cameras and wearable biometric devices to modify the classroom environment (e.g., through changes in images, lighting, and sound). The proposal for this project is to implement a real-time emotion detection system using portable Electroencephalography (EEG) that can be integrated into the interactive platform of Neurohumanities Lab.

   Our solution is both timely and fitting to address the outlined challenge. It revolves around developing a real-time prediction model for the previously mentioned emotions (admiration, love, hate, desire, joy, and sadness) using brain signals as input. This model is seamlessly integrated into the Neurohumanities Lab's interactive platform. This integration makes the platform an ideal tool for educational innovation, offering students an immersive experience. As they interact within this enriched environment, they can explore and deepen their understanding of Humanities in a classroom setting, which traditionally might not have had such technological engagement. At the same time, students and teachers can understand emotion generation during such experiences and analyze them in context.
   
   Central to our solution is the use of EEG signals. When properly processed, these signals reveal features pivotal for classifying the target emotions. The use of EEG in emotion recognition is not novel; its efficacy has been demonstrated in previous studies~\citep{Islam, Valenzi}. Many studies suggest that EEG signals provide enough information for detecting human emotions with feature-based classification methods~\citep{Valenzi}. Others have shown that emotional processing in the brain can be seen from the asymmetry in the brain activity recorded by EEG~\citep{Brown}.

   Various models have been proposed in the intriguing journey to understand and classify human emotions. One of the notable ones is the Circumplex 2D model put forward by James Russel~\citep{Rusell}. This innovative model utilizes a two-dimensional approach, mapping emotions based on Valence and Arousal. Valence measures the emotion's intrinsic appeal, determining whether it is perceived as positive or negative. On the other hand, Arousal gauges the level of excitement or intensity associated with the emotion. However, the quest for deeper understanding did not stop there. A subsequent, more intricate, 3D model known as Pleasure, Arousal, and Dominance (PAD), or Valence, Arousal Dominance (VAD), came to the fore~\citep{Islam}. This model broadened the horizon by incorporating these three elements. While Pleasure and Arousal are reminiscent of Russel's 2D model, adding Dominance provides additional insight. Dominance delves into the realm of control, assessing the extent to which an individual feels in command of, or subdued by, a particular emotion~\citep{Islam}. This addition elevates the complexity of the model, shedding light on the dynamic interplay between emotions and the sense of power or submission they instill.

   The EEG data acquisition process is characterized by several factors: the number of electrodes/channels, electrode placement system on the scalp (measurement of different brain regions), types of stimuli, sampling frequency, and the device used for signal acquisition. The International 10-20 electrode placement system is commonly used for emotion recognition using EEG~\citep{Islam}.
   
   The most fundamental and challenging task of recognizing human emotion is to find the most relevant features that vary with emotional state changes. The extracted EEG features for shallow and deep learning-based emotion recognition methods are the following: Time-domain features, which include statistical features such as mean, median, standard deviation, mode, variance, minimum, and maximum. The EEG frequency domain features usually contain more relevant information. The main methods are Power Spectral Density (PSD), Fast Fourier Transform (FFT), and the Short Time Fourier Transform (STFT)~\citep{Chaudhary, Lee}. The Wavelet transform method of analysis presents a good performance both in the time and frequency domain~\citep{Mohammadi} and can be classified into two types: Continuous Wavelet Transform~\citep{Bostanov} and Discrete Wavelet Transform~\citep{Islam}. The frequency domain approach was used for this work, focusing on PSD analysis. PSD analysis is a widely used technique for examining the power distribution of various frequency components in a signal and allows us to gain insight into the underlying frequency characteristics of the data. This approach enables the identification of prominent frequency bands or patterns that may indicate specific phenomena or attributes of the EEG signal. In addition, PSD analysis allows for quantifying the relative power contributions of different frequency components, providing valuable information for further analysis and interpretation of the mental states of the participants.

   EEG-based emotion recognition systems reported in the literature can be classified into two major groups: Deep machine learning-based and Shallow machine learning-based systems. The first one includes Convolutional Neural Networks (CNN), Deep Neural Networks (DNN), Deep Belief Networks (DBN), Recurrent Neural Networks (RNN), Bimodal Deep Auto Encoder (BDAE), Voting Ensembles (VEn), as classifiers. On the other hand, the second one includes Support Vector Machine (SVM), k Nearest Neighbor (kNN), Random Forest (RF), Decision Tree (DT), Multi-Layer Perceptron (MLP)~\citep{Islam}. Deep learning techniques are more effective than shallow learning-based algorithms among a wide range of algorithms. However, it may be noted that the SVM performs well in EEG-based emotion. Whenever portability and simplicity are not required, the multimodal data incorporating the other physiological signals (e.g., heart rate, skin conductance, among others) can significantly improve the performance of the emotion recognition system.
   
   In this work, an evaluation of different classification algorithms was implemented to obtain the most accurate model that classifies the desired emotions through VAD estimations in real time. In order to achieve this, a feature extraction, feature selection, and model evaluation process was followed. The length of time windows for the real-time estimation was also included as a parameter when evaluating the models. 

   Given the aforementioned background, this work describes a real-time emotion detection algorithm based on VAD estimations for the prediction of the six Descartes' passions. Section~\ref{sec:materials} presents the details about the dataset used during this work; Section~\ref{sec:methods} shows a detailed description of the data analysis, feature extraction and selection, and model performance evaluation; Section~\ref{sec:results} presents the results of the implementation and a detailed analysis of the variables and models, as well as the specifics of the best predictive model. Finally, section~\ref{sec:discussion} presents a discussion of the obtained results. 

\section{Materials} \label{sec:materials}

 \subsection{Datasets}

   A review on EEG-based emotion recognition algorithms using deep and shallow learning techniques is presented in~\citep{Islam}, analyzing 41 papers on this topic. Within those articles, 85\% use publicly open datasets; in the rest 15\%, a self-generated dataset was preferred. Among the 85\% articles using publicly available datasets, 61\%, 7\%, 2\%, and 15\% of the articles used: A Database for Emotion Analysis using Physiological Signals (DAEP)~\citep{Koelstra}, The Shanghai Jiao Tong University (SJTU) Emotion EEG Dataset (SEED)~\citep{Zheng2015}, MAHNOB~\citep{Soleymani2012} and other datasets, respectively. About 26\% used the images as stimuli, 23.8\% used video, 17.5\% used audio, 22.2\% used the existing dataset comprising a combination of physiological and emotional data~\citep{Alarcao}. The rest of the 10.5\% works exploited the emotional data related to games, live performances, or life events. Among these works, different researchers used a diverse range of frequency band-pass filters, and among them, the 4-45 Hz is predominantly used~\citep{Lakashmi}. 

 \subsection{DEAP Dataset}

    One of the main areas where Human-Computer Interfaces (HCI) are deficient is in the field of emotional intelligence. Most HCI systems are unable to interpret information derived from human emotional states and use it to prompt appropriate actions. With this in mind, the article by~\citep{Koelstra} describes a multimodal dataset that aids in analyzing human affective states. With this objective, the experiment was divided into two parts.
    
    The chosen dataset used to train the model for this project was the DEAP dataset~\citep{Koelstra}. The first part consists of a self-assessment where 16 subjects observed  120 one-minute videos and rated the three variables, Valence, Arousal, and Dominance, on a discrete (1-9) scale. The participants self-rated this scale using the Self Assessment Mannequins (SAM)~\citep{Bradley}. These three planes can be used to quantitatively describe emotion with (1) Arousal, ranging from inactive (uninterested) to active (excited); (2) Dominance, either feeling weak (without control) or empowered (in control); and (3) Valence, ranging from unpleasant to pleasant, where sadness or stress are considered unpleasant and happiness or excitement are considered as pleasant~\citep{Koelstra}.
    
     From 120 videos used, 60 were chosen semi-automatic using the "Last.fm" music enthusiast website, which allows users to assign tags to songs and retrieve songs assigned to those tags. In order to select these songs, a list of emotional keywords, as well as inflections and keywords, was used to generate a list of 304 keywords. For each of these keywords, a corresponding tag was searched in the "Last.fm" database, and the ten songs that were most often labeled with this particular tag were selected. This criterion yielded a total of 1084 songs (from which 60 of them were selected for the experiment). In order to choose these 60 songs, the valence-arousal space was divided into four quadrants, and 15 songs were manually selected for each quadrant. The quadrants are Low Arousal/Low Valence (LALV), Low Arousal/High Valence (LAHV), High Arousal/Low Valence (HALV) and High Arousal/High Valence (HAHV)~\citep{Koelstra}. In addition to those 60 videos, another 60 were manually selected (15 videos for each quadrant). 
     
    For each of the selected (120) videos, a 1-minute segment was extracted for the experiment. Subjects were asked to watch each video and provide a VAD rating. Based on the subjective rating obtained from the volunteers, 40 videos were selected out of the original 120 videos, where videos with the strongest ratings and smallest variance were selected for the second part of the experiment~\citep{Koelstra}.
    
    The second part of the experiment consists of 32 subjects who watched 40 videos in a laboratory environment with controlled illumination and rated through a self-assessment the familiarity of the video on a discrete scale of 1-5 and the liking, arousal, valence, and dominance on a continuous scale of 1-9. While the volunteers were watching the videos, EEG and peripheral physiological signals were recorded, and face video was recorded for 22 participants. Peripheral physiological signals included in this experiment are Galvanic Skin Response (GSR) and Photoplethysmography (PPG). 
    
    The database shows an in-depth analysis of the correlates between the EEG signals from the subjects and the subjective ratings given to each video in order to be able to propose a new method for stimuli selection for emotion characterization, providing a statistical analysis of the data obtained~\citep{Koelstra}.

\subsection{NHLab Functionality}

    The NeuroHumanities (NH) Lab's immersive platform integrates three primary functionalities: Real-time emotion detection, movement detection, and brain synchronization, which form the foundation of the proposed interactive platform.

    The real-time emotion detection functionality aims to monitor and identify the emotions of individuals within the platform's space. This capability allows for real-time environmental adjustments (colored lights, sounds, music) based on detected emotions and physiological signals, ensuring a tailored experience that aligns with the user's current emotional state. The movement detection functionality provides an interactive dimension to the platform. It enables individuals to interface with the environment using bodily movements and carry out interactive tasks such as painting over a projected screen, selecting words from a projection, and generating changes on images using facial expression recognition. This interaction facilitates active participation and increases user engagement, fostering a more effective learning process. The brain synchronization function seeks to analyze the synchrony between the brain activity of two users. By monitoring the synchronization between EEG channels of both and comparing it with their concurrent behaviors, insights into neural activity related to different aspects of social interactions can be obtained.

    By merging these three functionalities, the NH Lab's immersive platform offers a comprehensive educational experience that incorporates real-time emotion monitoring, active user movement interaction, and insights into brain synchronization.

 \subsection{Emotion Classification Model} \label{subsec:emotionclass}
 
\begin{figure}[ht]
     \centering
     \includegraphics[trim = 0 0 0 70, clip, width=\linewidth]{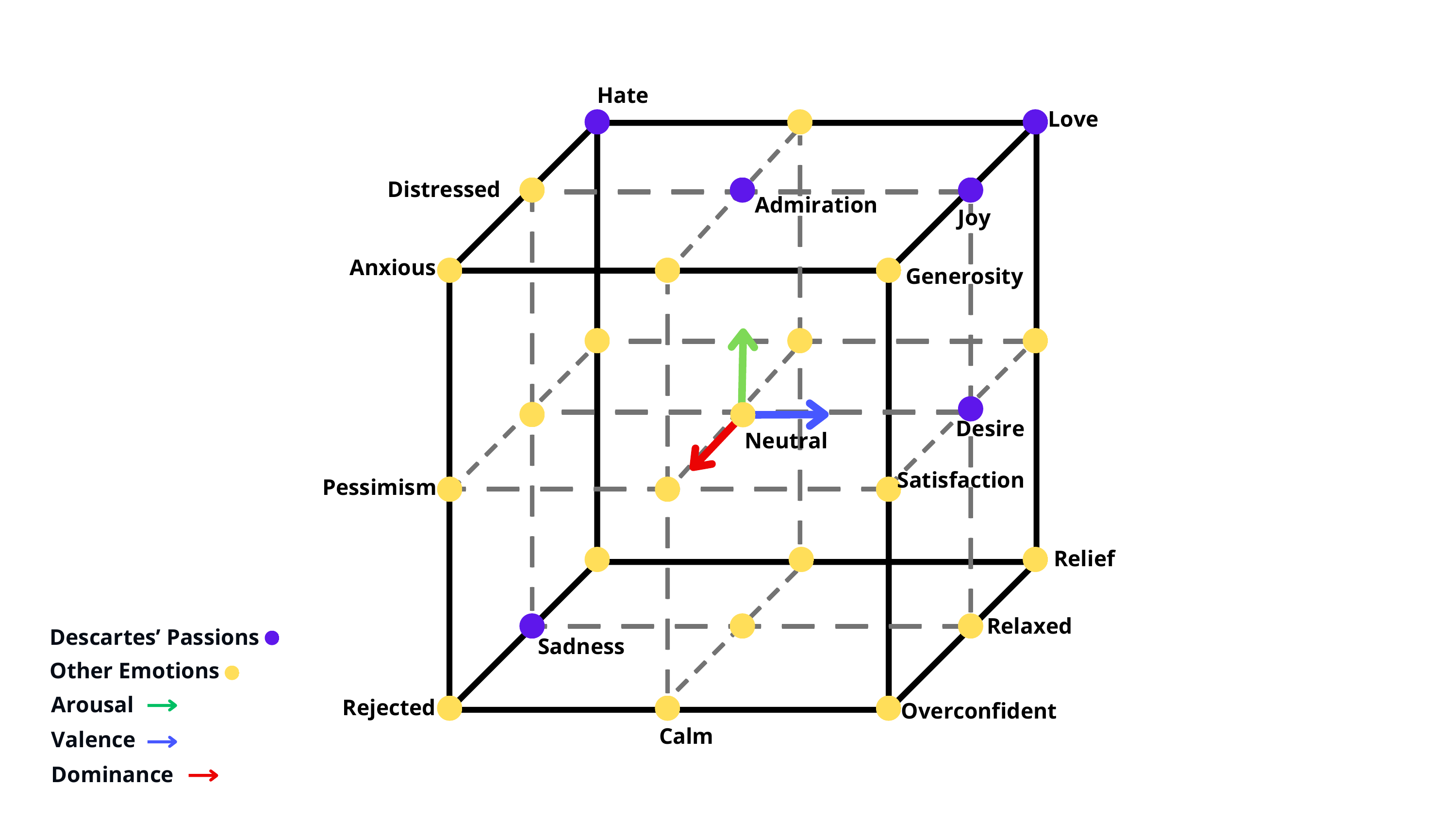}
     \caption{Three-dimensional VAD emotion model based on the work by~\citep{Islam}. Each VAD value had their unique direction and color: Arousal is green and at the y-axis (upwards and downwards), Valence is blue and at the x-axis (leftwards and rightwards) and dominance is red and at the z-axis (in and out). Additionally, Descartes' emotions are colored in purple (Hate, Love, Admiration, Joy, Desire, Sadness), while others are are colored in yellow.}
     \label{fig:cube}
    \end{figure}

    The classification model used in this study is based on a 3D model of emotion, which utilizes the VAD framework, shown at Figure~\ref{fig:cube}. This 3D model represents a three-coordinate system, with each coordinate corresponding to one of the VAD labels.

    In order to obtain emotions based on the VAD values, a new classification framework was created. Values between 1 and 3.6 were considered low and represented by -1 in the 3D model. Values between 3.7 and 6.3 were considered as medium and represented by 0. Values between 6.4 and 9 were considered high and represented by 1 in the 3D model.
    
    Coordinates were then obtained based on the relationship between the three classes, where each class is associated with a specific emotion. The 3D emotion model used in this study is based on the six different passions proposed by Descartes, including admiration, love, hate, desire, joy, and sadness. The 3D model of emotion used is based on~\citep{Islam}, with added passions of Descartes. The emotions highlighted in Table~\ref{tab:EmotionMap} are selected for classification in this work. This approach allows for the classification and representation of emotions within the 3D model, providing a framework for understanding and analyzing the participants' emotional responses.

\begin{table}[ht]
\centering
\caption{\label{tab:EmotionMap} Relation between the emotion and its coordinate system based on the 3D model at Figure~\ref{fig:cube} Highlighted emotions are Descartes’ passions.}

\begin{tabular}{llll}
\footnotesize
Arousal & Valence & Dominance & Emotion \\
\hline
0 & 0 & 0 & Neutral\\
0 & 0 & 1 & Other\\
0 & 0 & -1 & Other\\
0 & 1 & 0 & \textbf{Desire}\\
0 & 1 & 1 & Other\\
0 & 1 & 1 & Other\\
0 & 1 & -1 & Satisfaction\\
0 & -1 & 0 & Other\\
0 & -1 & 1 & Pessimism\\
0 & -1 & -1 & Other\\
1 & 0 & 0 & \textbf{Admiration}\\
1 & 0 & 1 & Other\\
1 & 0 & -1 & Other\\
1 & 1 & 0 & \textbf{Joy}\\
1 & 1 & 1 & Generosity\\
1 & 1 & -1 & \textbf{Love}\\
1 & -1 & 0 & Distressed\\
1 & -1 & 1 & Anxious\\
1 & -1 & -1 & \textbf{Hate}\\
-1 & 0 & 0 & Other\\
-1 & 0 & 1 & Calm\\
-1 & 0 & -1 & Other\\
-1 & 1 & 0 & Relaxed\\
-1 & 1 & 1 & Overconfident\\
-1 & 1 & -1 & Relief\\
-1 & -1 & 0 & \textbf{Sadness}\\
-1 & -1 & 1 & Rejected\\
-1 & -1 & -1 & Other\\
\end{tabular}\par
\centering
\end{table}

\section{Methods} \label{sec:methods}

\subsection{Data Preparation}
\label{sec:dataprep}
A pre-processed subset of data was used from the aforementioned DEAP Dataset to obtain the desired emotion detection model. The pre-processing consists of three main steps: (1) downsampling the data from 512 Hz to 128 Hz, (2) applying a band-pass filter between 0.4 Hz and 45 Hz, and (3) averaging the data to the common reference. These files were then combined, and the VAD values were extracted after applying the continuous to discrete transformation explained in Section~\ref{subsec:emotionclass}.

\begin{figure}[!ht]
\begin{center}
    \centering
    \includegraphics[width=0.75\linewidth]{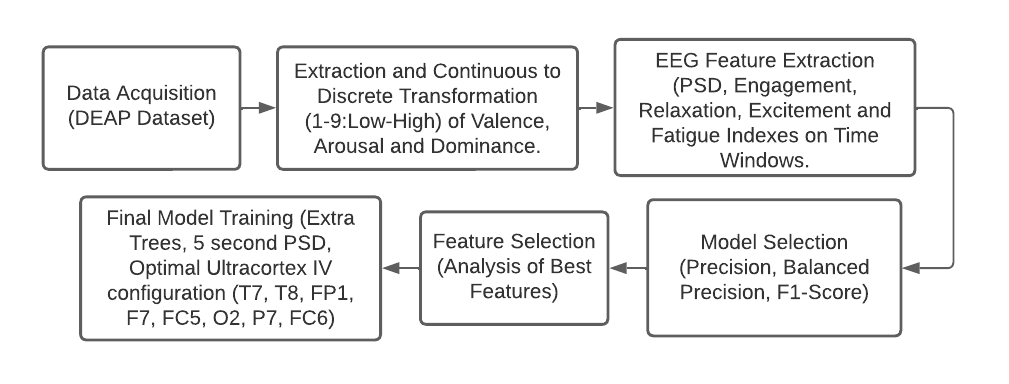}
    \caption{Flowchart of the Model Training Process, including feature extraction, selection, model performance evaluation, and best model selection.}
    \label{fig:mlfch}
\end{center}
\end{figure}

\subsection{Preliminary Model Testing}

In the initial stages of the project, a preliminary analysis was performed to obtain a suitable machine-learning model for extracting valuable insights from the EEG data. A comparative study involved three well-established algorithms: the RF Classifier, XGBoost, and the SVM. These models were selected based on their known efficacy in handling high-dimensional data. For the preliminary analysis, data with all the available channels was used to train these models. Evaluation metrics, including accuracy, sensitivity, and specificity, were then used to measure the performance of each model. This foundational step was deemed critical for guiding the subsequent feature and channel selection procedures and validating them.

\subsection{Feature Engineering}
\label{sec:featureeng}
\subsubsection{PSD Estimation}
\label{sec:PSD}

The pre-processed EEG data was segmented into 5-second time windows to capture transient and evolving neural dynamics over time. FFT was then applied to each segment, transforming the raw time-domain EEG data into the frequency domain to obtain PSD using the FFTProcessing function created by~\citet{tongdaxu}. 

PSD values were segmented into five distinct frequency bands: Delta ($\delta$) [0.5-4 Hz], theta ($\theta$) [4-8 Hz], alpha ($\alpha$) [8-12 Hz], beta ($\beta$) [12-30 Hz], and gamma ($\gamma$) [30-45 Hz]. 

Selecting an optimal time window length for practical real-time EEG analysis remains critical. A balance between precision and computational speed is essential: while longer windows often provide increased resolution and potential model accuracy, they may compromise processing speed and real-time responsiveness. On the other hand, shorter windows can facilitate rapid processing but might reduce precision. In order to address this balance, the accuracy of VAD prediction was evaluated across different time window lengths and identified an optimal window length. This strategic decision underpins the reliability and efficiency of the real-time implementation.

\subsubsection{Frequency Band Ratios}
\label{sec:indexes}

To aid with the models' performance, different frequency band ratios were implemented for each channel, 

\textbf{Relaxation Index ($\frac{\theta}{\delta}$)}: This ratio accentuates the interplay between the theta waves, linked with drowsiness or light meditation states, and delta waves, associated with deep sleep and unconscious processes~\citep{Borges2010}. An elevated Theta/Delta ratio often suggests a state of relaxation without lapsing into deep unconsciousness, making it a valuable marker for assessing relaxation in awake individuals.

\textbf{Excitement Index ($\frac{\beta}{\alpha}$)}: This ratio is served as an indicator of attention and engagement. It is suggested by a higher ratio that individuals are more alert and attentive, which can be interpreted as excitement or heightened interest. In the context of the research paper, the efficiency of advertisements at a population level was predicted using this ratio, indicating that ads evoking higher engagement, as measured by the beta/alpha ratio, are deemed more effective~\citep{Kislov2023}.

\textbf{Mental Fatigue Index ($\frac{\alpha}{\theta}$)}: By examining the Alpha/Theta ratio, signs of mental weariness can be observed. A predominant alpha wave activity in relation to theta can signify a relaxed or idling state of the brain, which, especially during tasks requiring sustained attention, can indicate cognitive fatigue~\citep{ijerph182211891}.

\textbf{Engagement Index ($\frac{\beta}{\theta+\alpha}$)}: In order to represent the balance of active cognitive processing versus a more passive state, this ratio is particularly crucial in contexts where the depth of cognitive immersion or focus is under scrutiny. A high value suggests robust cognitive engagement or alertness, paramount in activities demanding continuous mental effort~\citep{Ismail2020}.

\subsection{Feature Selection}

\subsubsection{Channel Selection using PCA and RF} 
\label{sec:met8}

 Among the crucial features of the real-time model, it is noteworthy that the evaluated dataset includes data from 32 EEG channels; however, the proposed real-time algorithm is intended to be integrated into an 8-channel, dry-electrode OpenBCI system (OpenBCI, New York, NY, USA) for a highly-portable, practical implementation~\citep{Lakhan, Zhou}. Many studies have used only 8 channels to obtain EEG signals \citep{Brown, mikhail, Valenzi, Zhou}. This 8-channel system allows the re-configuring of different positions of electrodes around the scalp. 
 
The channel selection was conducted to assess which are the best 8 channels to use by the OpenBCI for the proposed emotion detection algorithm. Following this idea, channels were grouped into their respective lobes and joined their data via dimensionality reduction techniques; then, three RF models were fitted (one for each emotion component) in order to assess lobe importance with respect to a series of frequency bands.

Feature importance becomes quite challenging given the high dimensionality of source features ($32$ channels $\cdot$ $5$ frequency bands = $160$), and the vast amount of total samples ($593,920$; $58$ seconds of a video clip through a $0.125$ second moving window, $58/0.125 = 464$ samples for each $40$ videos and $32$ subjects, $464 \cdot 40 \cdot 32 = 593,920$). Regression models such as RF would be not only slow due to high dimensionality, but slight differentiation between features' importance from one to another would be absent due to using normalized relative importance, which gives each feature an importance such that the sum of importance is equal to 1.

PCA was used to reduce the dimensionality of the data; it was applied to source features regarding their channel anatomical location, separated within the Frontal, Temporal, Parietal, Occipital, Central, and Central-Parietal lobes, for each frequency band, and subjects' data. Table~\ref{tab:pca} shows which EEG channels were assigned to which lobe.

\begin{table}[htbp]
\centering
\caption{Brain lobes and channels are assigned to each lobe regarding brain anatomy; although some lobes have more channels than others, PCA always reduces the dimensionality to $k$ components.}
\begin{center}
\begin{tabular}{ll}
Lobe & Channels \\
\hline
Frontal & Fp1, Fp2, F3, F4, F7, F8, Fz, AF3, AF4 \\
Temporal & T7, T8 \\
Parietal & P3, P4, P7, P8, Pz, PO3, PO4 \\
Occipital & O1, O2, Oz \\
Central & FC5, FC1, C3, C4, FC2, FC6, Cz \\
Central-Parietal & CP5, CP1, CP2, CP6 \\
\end{tabular}
\label{tab:pca}
\end{center}
\end{table}

The objective of applying PCA to the dataset is to gather insights about the brain region whose features are most linearly correlated to the target features, which would suggest a strong relation. The calculation of the PCA first consists of the Singular Value Decomposition (SVD) technique, which decomposes matrix $X$ into the matrix multiplication of three matrices $U$ $\Sigma$ $V^T$, as in Equation~\ref{eq:SVD}.

\begin{equation}
    X = U \cdot \Sigma \cdot V^T
    \label{eq:SVD}
\end{equation}

$V$ has the unit vectors of the $k$ components, represented in Equation~\ref{eq:V}.

\begin{equation}
    V = 
    \begin{pmatrix}
        \vert & \vert & & \vert \\
        c_1 & c_2 & \cdots & c_k \\
        \vert & \vert & & \vert
    \end{pmatrix}
    \label{eq:V}
\end{equation}

$V$ has $n \times k$ dimensionality, although we only require the first component in order to compress all the channels' vectors on each lobe to a single vector. So for each $V$ matrix calculated, when considering the first component $W$, which is the eigenvector corresponding to the largest eigenvalue of the covariance matrix, for each lobe $l$ and frequency band $b$ with respect to a subject $s$, the whole $X$ data (consisting of $n$ samples by $m$ channels), was projected into a $Z$ vector (consisting of $n$ samples by $1$ dimension), thus leaving a unique vector of values for each combination of $l$ and $b$ on a subject $s$. The computation can be seen in Equation~\ref{eq:transform}.

\begin{equation}
    Z_{s, l, b} = X_{s, l, b} \cdot W_{s, l, b}
    \label{eq:transform}
\end{equation}

Pearson correlation coefficient is calculated for each subject $s$ as in Eq.~\ref{eq:pearson}; afterwards, an average across $S$ subjects was calculated; variance between each subject's data was thus reduced as if PCA would be calculated on all complete data, features' domain between subjects would make PCA unstable, as the technique assumes that the dataset is centered around the origin; where $z_i$ represents each sample for each PCA $l$ lobe and $b$ frequency band, while $y_i$ represents each sample for each $e$ emotion (VAD) component. Thus, a single mean absolute correlation coefficient is calculated for each source feature, averaged among each subject's data. This coefficient was further used in order to determine which channels are the most linearly correlated to the target emotion VAD, hence reducing the dimensionality of the whole dataset when sub-setting the best lobes on a particular frequency band for each emotion component.

\begin{equation}
    r(l, b, e) = \frac{1}{S} \sum_{s=1} ^{S} \vert\frac{\sum_{i=1}^n (z_{i, s, l, b}-\overline{z}_{s, l, b})(y_{i, s, e}-\overline{y}_{s, e})}{\sqrt{\sum_{i=1}^n(z_{i, s, l, b}-\overline{z}_{s, l, b})^2\sum_{i=1}^n(y_{i, s, e}-\overline{y}_{s, e})^2}}\vert
    \label{eq:pearson}
\end{equation}

Once the most relevant lobes were found, an RF regressor was used to determine feature importance, thus obtaining the best $8$ channels using feature selection. The RF model from Sklearn, an ML package from Python, contains the Gini importance (GI) metric, which is the normalized total reduction of a given criterion brought by each feature. On classification, it represents the number of splits in a decision tree that used that feature within the RF ensemble~\citep{CandelaLeal2021}; while on regression, it uses the Mean Squared Error (MSE) criterion, which follows the Euclidean norm, thus reducing the Euclidean distance between two points in a given vector space~\citep{CandelaLeal2022} and giving highest importance to the feature that reduced most variance.

The MSE criterion is shown in Equation~\ref{eq:MSE}, which is essential when calculating GI in RF regression; the reduction of variance is calculated by minimizing the squared difference between target feature $y$ and predicted target feature $\hat{y}$, based on RF model. The reduction of this criterion, at most, by including a specific channel $c$ as a feature would then have the highest GI normalized across all $C$ channels.
    
\begin{equation}
    MSE(c, e) = \frac{1}{N} \sum_{i=1} ^{N} (y_{i, e} - \hat{y}_{i, c, e})^2
    \label{eq:MSE}
\end{equation}

GI was obtained for each subject's data $s$, then averaged across the total number of subjects $S$. Although RF is randomly initialized, to generalize better, $I$ iterations were carried out in order to ensure that a random initialization would not benefit a specific feature (in which $I=10$). Therefore, for each subject $s$, $I$ RF models with different random initializations were run, and GI was averaged to obtain GI for a specific subject. Furthermore, each subject's GIs were averaged to obtain a global importance based on all subjects. This aforementioned calculation is shown in Equation~\ref{eq:GIs}.

\begin{equation}
    GI(c, e) = \frac{1}{S} \frac{1}{I} \sum_{s=1} ^{S} \sum_{i=1} ^{I} GI(c, e)_{i, s}
    \label{eq:GIs}
\end{equation}

Considering only the best band and lobe combination, the criterion is computed for $32$ channels $c$ ($C=32$); hence, the normalized feature selection criterion would be capable of detecting slight importance changes between features. Three RF models, one for each emotion $e$ component (VAD), would be fitted; thus, each emotion component would have their best set of features, in which Equation~\ref{eq:MSEemo} must be satisfied. 
    
\begin{equation}
    \sum_{c=1} ^{C} GI(c, e) = 1
    \label{eq:MSEemo}
\end{equation}

Given that the three GI would have the same domain and the same number of features, an Emotion Importance Index (EII) was calculated in order to evaluate the overall importance of each feature to the process of predicting $E$ emotions, as in Equation~\ref{eq:EII}, hence proposing a more holistic approach on the feature selection process of selecting the best $8$ EEG channels.

\begin{equation}
    EII(c) = \frac{1}{E} \sum_{e=1} ^{E} GI(c, e)
    \label{eq:EII}
\end{equation}

\subsection{Model evaluation}

\subsubsection{Model Selection}
\label{sec:modelselec}
    
A range of metrics was adopted to identify the optimal classifier model. Recognizing that a single metric might not fully capture a model's effectiveness, especially with varied data distributions, three metrics were employed: accuracy, balanced accuracy, and F1-score. These metrics were selected to offer a comprehensive understanding of model performance and to ensure a reliable choice was made. Metrics were calculated as in~\citep{AguilarHerrera2021}, where True Positives (TP), True Negatives (TN), False Positives (FP) or type-I errors, and False Negatives (FN) or type-II errors are used to build up these metrics.

\textbf{Accuracy:} In the model selection process, accuracy was utilized as a primary metric. Defined by the equation:

\begin{equation}
    \text{Accuracy} = \frac{TP + TN}{TP + TN + FP + FN} 
    \label{eq:Acc}
\end{equation}

Accuracy (Equation~\ref{eq:Acc}) represents the ratio of correctly predicted observations to the total observations. This measure assessed the overall correctness of the model's predictions.

\textbf{Balanced Accuracy (BA):} Given the potential pitfalls of using accuracy alone, especially in imbalanced datasets, balanced accuracy was also employed. It is computed as the average of the true positive rate, or sensitivity (Equation~\ref{eq:Sens}) and the true negative rate, or specificity (Equation~\ref{eq:Spec}), given by:

\begin{equation}
    \text{Balanced Accuracy} = \frac{\text{Sensitivity} + \text{Specificity}}{2}
    \label{eq:Bacc}
\end{equation}

Where

\begin{equation}
    \text{Sensitivity} = \frac{TP}{TP + FN} 
    \label{eq:Sens}
\end{equation}

And

\begin{equation}
    \text{Specificity} = \frac{TN}{TN + FP} 
    \label{eq:Spec}
\end{equation}

This metric gained a more nuanced understanding of the model's performance on both the positive and negative classes.

\textbf{F1-Score:} For further depth in evaluation, the F1-score was incorporated. Represented by the equation:

\begin{equation}
    \text{F1-Score} = 2 * \frac{\text{Sensitivity} \cdot \text{Specificity}}{\text{Sensitivity} + \text{Specificity}}
    \label{eq:F1sc}
\end{equation}

By employing these metrics, a comprehensive understanding of the performance of different classifier models was ensured, allowing for a more informed model selection to be made. For each of the VAD classes (Low, Medium, and High), a total of 5 different classifier models were trained. Including three tree-based classification models: Extra-Trees (ET), Random Forest (RF), XGBoost (XGB); as well as other models such as k-Nearest Neighbor (kNN), Support Vector Classifier (SVC).

\subsubsection{Feature Importance Selection}

Once the model was selected, its performance was evaluated over a range of feature subsets, specifically between 20 and 40 features. The optimal subset was identified based on the best performance metrics. The feature importance function from scikit-learn, applied using the ExtraTrees model, was utilized for this assessment.

In tree-based models such as Extra-Trees, a feature's importance is determined by the frequency and depth of its use for data splits across all trees~\citep{Olivas2021}. A feature frequently used and closer to the tree roots is considered more crucial. The importance of a feature is typically calculated by averaging the decrease in impurity, often quantified using the Gini criterion, across all nodes where the feature facilitates a split~\citep{OlivasOM2021}. By aggregating over the ensemble of trees, more robust and less biased feature importances are typically achieved.

\subsection{Real-time implementation}

\begin{figure}[!ht]
\begin{center}
    \centering
    \includegraphics[width=\linewidth]{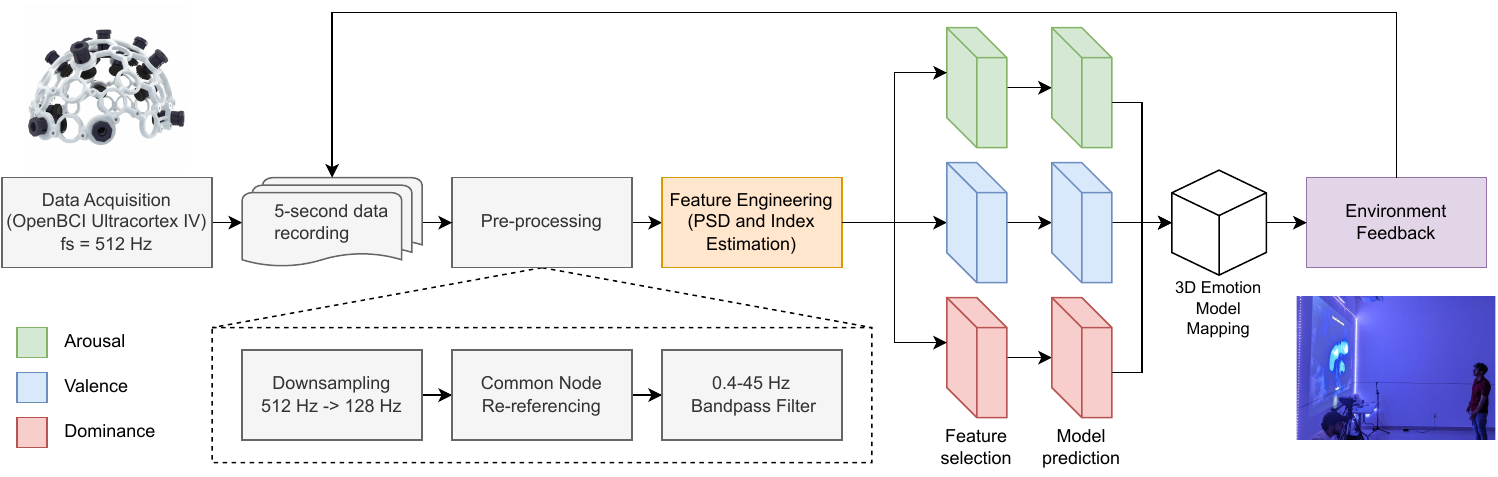}
    \caption{Flowchart of the real-time emotion detection implementation.}
    \label{fig:rtfch}
\end{center}
\end{figure}

In this work's final phase, the real-time application was developed. Emotion detection was achieved by integrating insights derived from three distinct Machine Learning (ML) models. This methodology was further enriched by including the three-dimensional emotion model proposed by~\citep{Islam}. Figure~\ref{fig:rtfch} shows the real-time pipeline implemented for the EEG-based VAD estimation and thus detecting of Descartes' passions.

The pipeline consists of retrieving EEG signals from an 8-channel OpenBCI Ultracortex IV EEG helmet using a Cyton board. Further, using 5-second data and pre-processing it according to Section~\ref{sec:dataprep}, features are created according to Section~\ref{sec:featureeng}, further subsetting the best features according to each VAD model, as well as their respective prediction. Finally, the user received feedback based on the VAD predictions and the 3D emotion model mapping described in Figure~\ref{fig:cube}. The OpenBCI Ultracortex IV EEG helmet has a total of 35 possible node locations, with the default version being FP1, FP2, C3, C4, P7, P8, O1, and O2. However, these channels were used in the initial iteration of the project; these would be further replaced with the optimal emotion recognition channels regarding the channel selection results in Section~\ref{sec:chansel}.

\begin{figure}[ht]
     \centering
     \includegraphics[width=0.85\linewidth]{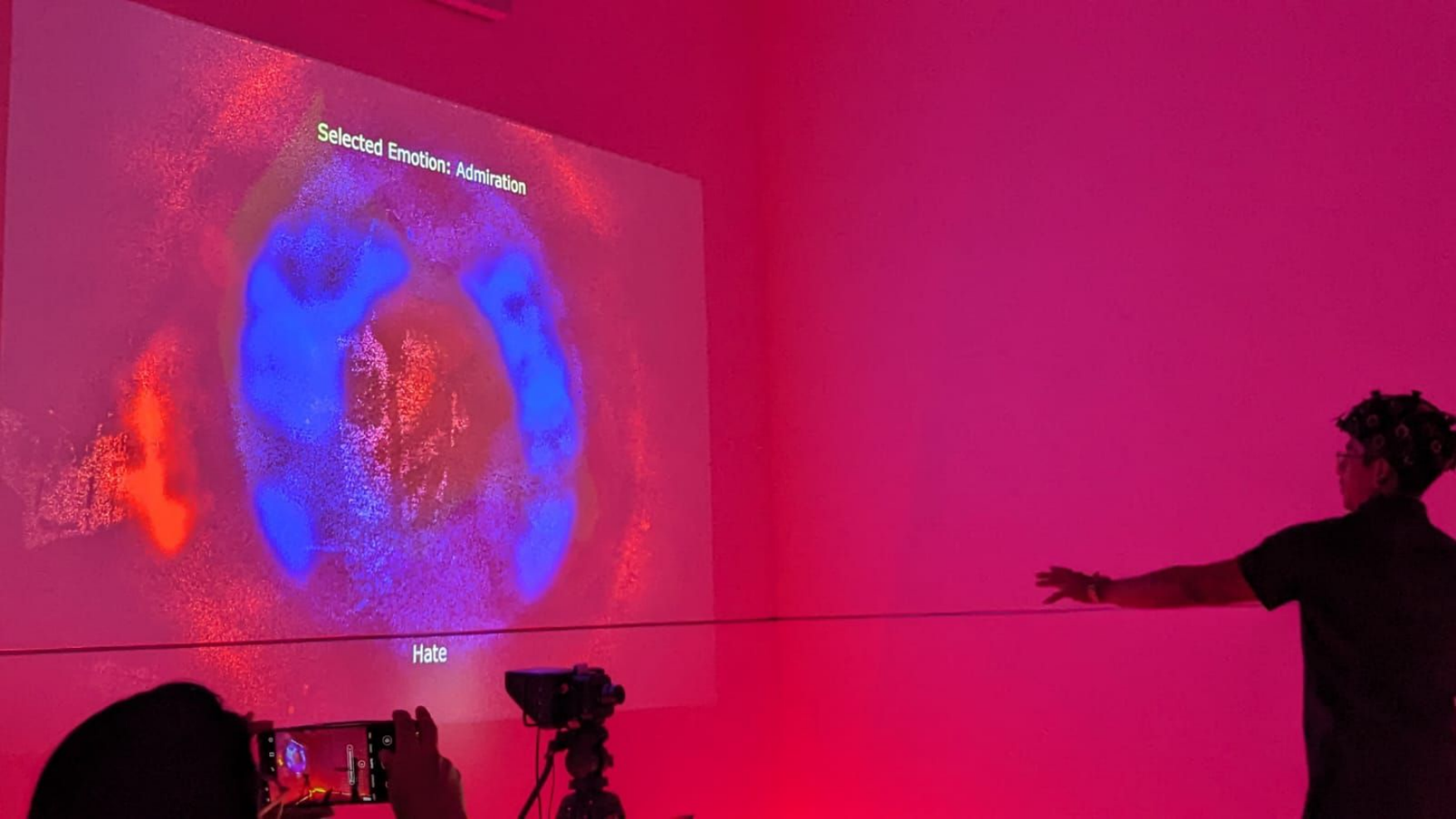}
     \caption{Testing the real-time application of the algorithm at the NeuroHumanities Lab at Tecnologico de Monterrey, Campus Monterrey}
     \label{fig:NHLab}
\end{figure}

On the other hand, the real-time use of the NH Lab platform can be seen in Figure~\ref{fig:NHLab}. In this platform, the user wears the OpenBCI helmet, and the emotion detection model identifies in real-time one of the six Descartes' passions. Depending on the detected emotion, the lighting of the environment changes. A camera, and a motion tracking algorithm are used to detect the users' movements that allow the generation of a painting on the projected screen. The color palette of the visualization, as well as sound effects and music related to the movement, are different depending on the detected emotion. 

Moreover, the participants wear an Empatica E4 (Empatica Inc, Milano, IT), which measures Electrodermal Activity (EDA), Blood Volume Pulse (BVP), 3-axis acceleration, and temperature. Empatica E4 signals are also acquired in real time. Changes in these values are reflected in the system as well; for instance, the volume of specific sounds related to the EDA (electrostatic noise) and BVP (heartbeats) changes in accordance with increases/decreases of the estimated engagement (via EEG).

\section{Results and Discussion} \label{sec:results}

\subsection{Time Window Selection}

Using the standard OpenBCI channel configuration, PSD was obtained over various time window lengths, ranging from 1 to 10 seconds. In order to determine the optimal window length, the performance-to-time ratio was considered, especially since the intended application was in real-time scenarios~\citep{LozoyaSantos2022}. The results for time windows of 2, 4, 5, 8, and 10 seconds are presented in Table~\ref{tab:timewindow}.

\begin{table}[ht]
    \centering
    \begin{tabular}{ccc}
        Window length (s) & VAD & Accuracy (\%) \\
        \hline
        \multirow{3}{*}{2} & A & 81 \\
        & V & 80 \\
        & D & 83 \\
        \hline
        \multirow{3}{*}{4} & A & 83 \\
        & V & 86 \\
        & D & 84 \\
        \hline
        \multirow{3}{*}{5} & A & 88 \\
        & V & 88 \\
        & D & 88 \\
        \hline
        \multirow{3}{*}{8} & A & 88 \\
        & V & 88 \\
        & D & 88 \\
        \hline
        \multirow{3}{*}{10} & A & 87 \\
        & V & 90 \\
        & D & 88
    \end{tabular}
    \caption{Accuracy comparison of different time windows on the three VAD components prior to the feature selection process using a standard RF model.}
    \label{tab:timewindow}
\end{table}

From the results, the 5-second time window was selected for the final model since it provided an average accuracy of 88\%, which was also observed for 8-second windows. Moreover, no significant improvements were observed for the 10-second window. Therefore, given the positive accuracy and the fast implementation it entails, the 5-second window was selected.

\subsection{Channel Selection} \label{sec:chansel}
 
    According to the lobe regions defined in Table~\ref{tab:pca}, Sklearn library from Python was used to obtain the first component for each lobe and spectral band combination ($6$ lobes $\cdot$ $5$ spectral bands = $30$). This process significantly reduces the number of features analyzed, in contrast to the initial ($32$ channels $\cdot$ $5$ spectral bands = $160$). After the PCA was calculated for each subject,  the mean absolute Pearson correlation coefficient was obtained, as stated in Equation~\ref{eq:pearson}, which provided a unique importance value for each combination tested.

    Results of the PCA-found features' correlation are shown in Table~\ref{tab:pca_r}. A unique VAD value was calculated for each frequency b   and and lobe combination; the summation of the three coefficients was also calculated to evaluate overall feature importance. The highest linearly correlated features correspond to the $\gamma$ frequency band, as it has the best overall correlation among VAD for each lobe: Frontal with $0.4272$, Temporal with $0.4433$, Parietal with $0.4585$, Occipital with $0.4563$, Central with $0.4443$, Central-Parietal with $0.4251$; along with the best correlation coefficients for each emotion component: $\gamma_O$ for Arousal with $0.2011$, $\gamma_P$ for Valence with $0.1316$, $\gamma_T$ for Dominance with $0.1327$.

\begin{table}[htbp]

\centering
\caption{Mean absolute Pearson correlation coefficients between the emotion component and each lobe's first Principal Component across each subject. F = Frontal, T = Temporal, P = Parietal, O = Occipital, C = Central, CP = Central-Parietal. The highest values for Arousal, Valence, and Dominance are marked in bold, in addition to the highest sum of VAD coefficients for each lobe. Statistically significant linearly correlated for more than $95\%$ of the subjects ($31$ out of $32$ or $32$ out of $32$ subjects) are marked with a $^*$.}
\begin{center}
\begin{tabular}{ll|llll}
$f_{band}$ & Lobe & Arousal & Valence & Dominance & $\Sigma$ \\
\hline

\multirow{6}{*}{$\delta$}
& F & $0.0472$ & $0.0447$ & $0.0422$ & $0.1341$ \\
\cline{3-6}
& T & $0.0511$ & $0.0455$ & $0.0436$ & $0.1402$ \\
\cline{3-6}
& P & $0.05$ & $0.0435$ & $0.0445$ & $0.138$ \\
\cline{3-6}
& O & $0.0438$ & $0.0423$ & $0.0426$ & $0.1287$ \\
\cline{3-6}
& C & $0.0511$ & $0.043$ & $0.0447$ & $0.1388$ \\
\cline{3-6}
& CP & $0.0456$ & $0.0446$ & $0.0406$ & $0.1308$ \\
\hline

\multirow{6}{*}{$\theta$}
& F & $0.0877$ & $0.0795$ & $0.0778$ & $0.245$ \\
\cline{3-6}
& T & $0.0954$ & $0.0804$ & $0.0785$ & $0.2543$ \\
\cline{3-6}
& P & $0.0971$ & $0.083$ & $0.0833$ & $0.2634$ \\
\cline{3-6}
& O & $0.0946$ & $0.0762$ & $0.0794$ & $0.2502$ \\
\cline{3-6}
& C & $0.0945$ & $0.0783$ & $0.0812$ & $0.254$ \\
\cline{3-6}
& CP & $0.0839$ & $0.083$ & $0.0773$ & $0.2442$ \\
\hline

\multirow{6}{*}{$\alpha$}
& F & $0.1146^*$ & $0.0686$ & $0.0763$ & $0.2595$ \\
\cline{3-6}
& T & $0.1123$ & $0.073$ & $0.078$ & $0.2633$ \\
\cline{3-6}
& P & $0.1138$ & $0.087$ & $0.0845$ & $0.2853$ \\
\cline{3-6}
& O & $0.1233$ & $0.0883^*$ & $0.0844$ & $0.296$ \\
\cline{3-6}
& C & $0.1182^*$ & $0.0828^*$ & $0.0836$ & $0.2846$ \\
\cline{3-6}
& CP & $0.1108$ & $0.0815$ & $0.08$ & $0.2723$ \\
\hline

\multirow{6}{*}{$\beta$}
& F & $0.1642$ & $0.0989$ & $0.1065$ & $0.3696$ \\
\cline{3-6}
& T & $0.1684$ & $0.0988$ & $0.1183$ & $0.3855$ \\
\cline{3-6}
& P & $0.1894$ & $0.1113$ & $0.1184$ & $0.4191$ \\
\cline{3-6}
& O & $0.1797$ & $0.1072$ & $0.1094$ & $0.3963$ \\
\cline{3-6}
& C & $0.1788^*$ & $0.0985$ & $0.1207$ & $0.398$ \\
\cline{3-6}
& CP & $0.1696^*$ & $0.1083$ & $0.1084$ & $0.3863$ \\
\hline

\multirow{6}{*}{$\gamma$}
& F & $0.1868^*$ & $0.1228$ & $0.1176^*$ & \textbf{0.4272} \\
\cline{3-6}
& T & $0.188^*$ & $0.1226^*$ & \textbf{0.1327} & \textbf{0.4433} \\
\cline{3-6}
& P & $0.2005$ & \textbf{0.1316}$^*$ & $0.1264$ & \textbf{0.4585} \\
\cline{3-6}
& O & \textbf{0.2011} & $0.1295$ & $0.1257$ & \textbf{0.4563} \\
\cline{3-6}
& C & $0.1958$ & $0.1178$ & $0.1307$ & \textbf{0.4443} \\
\cline{3-6}
& CP & $0.1783^*$ & $0.1281$ & $0.1187$ & \textbf{0.4251} \\
 
\multicolumn{6}{l}{$^{*} p < 0.05$ for $> 95\%$ of the subjects} 
\end{tabular}
\label{tab:pca_r}
\end{center}
\end{table}

    There is a linear correlation between frequency bands and the sum of Pearson correlation coefficients; when frequency increases, this coefficient also increases. Considering the overall sum coefficients for each lobe, on each frequency band: $\delta$ has coefficients between $0.12-0.14$, $\theta$ has coefficients between $0.24-0.26$, $\alpha$ has coefficients between $0.25-0.29$, $\beta$ has coefficients between $0.36-0.41$, $\gamma$ has coefficients between $0.42-0.45$. Hence, based on those overall coefficients, there seem to be three clusters: $\delta$, $\theta$ and $\alpha$, $\beta$ and $\gamma$, with the lowest frequency bands being the least linearly correlated and the highest frequency bands being the most linearly correlated. These results are similar to the reported by other authors~\citep{MuLi2009, Martini2012, Yang2020}, who also determined that $\gamma$ bandpower in EEG is the most suitable for emotion classification.

    It is important to note that linear correlation does not directly mean more feature importance, as some features might be non-linearly correlated and still be critical features for the target feature prediction. However, for the first assessment and feature discrimination, the assumption of higher lineal correlation means higher feature importance is made. Furthermore, an RF regression model would be fitted on the best three lobes' channels in order to gather true feature importance with respect to the MDI criterion, which does not necessarily give importance to linearly correlated features with the target emotion.    

    Regarding statistically significant linearly correlated features for more than $95\%$ of the subjects on at least one of the emotion components (VAD), the higher frequency bands trend is displayed, as $4$ features correspond to the $\gamma$ band, $2$ to the $\beta$ band, and $3$ to the $\alpha$ band. Arousal and Valence have several statistically significant linearly correlated features ($\alpha_A$, $\alpha_C$, $\beta_C$, $\beta_{PC}$, $\gamma_F$, $\gamma_T$, $\gamma_{PC}$) and ($\alpha_O$, $\alpha_C$, $\gamma_T$, $\gamma_P$) respectively, although Dominance only has $\gamma_F$.

\begin{figure}[!ht]
\begin{center}
    \centering
    \includegraphics[width=0.9\linewidth]{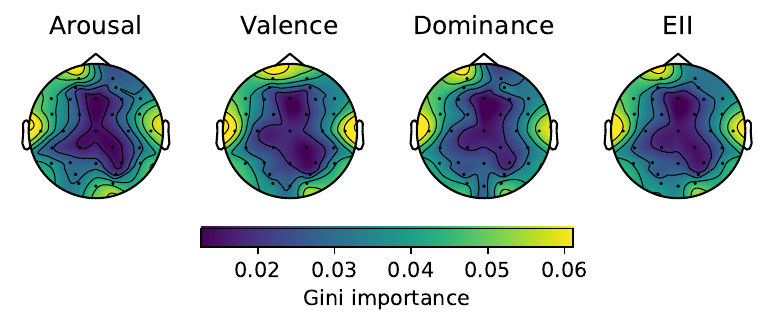}
    \caption{Average GI topoplot for each channel, considering only $\gamma$ frequency band (30-45 Hz) and fitting a separate RF model on Arousal, Valence, and Dominance. EII was calculated by averaging GIs across emotion components as described in Equation~\ref{eq:EII}.}
    \label{fig:gimp}
\end{center}
\end{figure}

    In order to assess which are the best $8$ channels to use in the OpenBCI helmet (available for real-time implementation), the best features $6$ lobes with the highest sum of VAD coefficients were considered ($\gamma_F$, $\gamma_T$, $\gamma_P$, $\gamma_O$, $\gamma_C$, $\gamma_{PC}$). Hence, all the lobes on the $\gamma$ frequency band would be a good choice, so each channel $c$ would have an assigned importance value (higher is better), calculated based on Equation~\ref{eq:GIs}, described in Section~\ref{sec:met8}. In order to visually understand the importance of each channel, a topoplot was created, illustrating a spatial map of the obtained GI values. The topoplot is shown in Figure~\ref{fig:gimp}.

    Overall, Arousal, Valence, and Dominance topoplots are similar, as the channels at the center of the brain are not as important as the ones at the exterior part of the brain; this pattern is also shared on EII, which displays the average patterns found on each of other three emotion component's plot. Focusing on EII's topoplot better generalizes overall feature importance on determining which channel is more related to VAD values, and thus it is more useful when identifying emotions.
    
    The coefficients are also presented in Table~\ref{tab:coefgamma}. Both T7 and T8 have the highest GI importance according to EII ($0.0586$, $0.0557$), which are the only channels from the temporal lobe, following Fp1 ($0.0519$), and F7 ($0.0442$) from the frontal lobe. These results are similar to the reported by~\citep{Zhang2016, Wang2019}, where temporal and frontal regions had the highest importance for emotion recognition.

\begin{table}[htbp]
\centering
\caption{GI for each 32 EEG DAEP channel at their $\gamma$ bandpower, ordered from highest to lowest importance according to EII. GIs from this table were used in order to create the topoplot at Figure~\ref{fig:gimp}.}
\begin{center}
\begin{tabular}{lllll|lllll}
Channel & A & V & D & EII & Channel & A & V & D & EII \\
\hline
T7 & $0.0561$ & $0.061$ & $0.0586$ & $0.0586$ & C4 & $0.0284$ & $0.0302$ & $0.0289$ & $0.0292$ \\
T8 & $0.0527$ & $0.0611$ & $0.0534$ & $0.0557$ & CP6 & $0.0292$ & $0.0271$ & $0.0299$ & $0.0287$ \\
Fp1 & $0.0495$ & $0.0523$ & $0.0539$ & $0.0519$ & P3 & $0.0292$ & $0.0301$ & $0.0268$ & $0.0287$ \\
F7 & $0.0439$ & $0.0435$ & $0.0451$ & $0.0442$ & CP5 & $0.0248$ & $0.0294$ & $0.0295$ & $0.0279$ \\
FC5 & $0.0425$ & $0.0433$ & $0.0444$ & $0.0434$ & PO3 & $0.0291$ & $0.027$ & $0.0256$ & $0.0272$ \\
O2 & $0.0442$ & $0.0429$ & $0.0426$ & $0.0432$ & C3 & $0.0311$ & $0.022$ & $0.0271$ & $0.0267$ \\
P7 & $0.0374$ & $0.0419$ & $0.0371$ & $0.0388$ & FC1 & $0.0242$ & $0.0239$ & $0.0283$ & $0.0255$ \\
FC6 & $0.0439$ & $0.0364$ & $0.0359$ & $0.0387$ & Pz & $0.0273$ & $0.0195$ & $0.0265$ & $0.0244$ \\
P8 & $0.0389$ & $0.0382$ & $0.0335$ & $0.0369$ & F4 & $0.024$ & $0.026$ & $0.0216$ & $0.0239$ \\
O1 & $0.0333$ & $0.0334$ & $0.0385$ & $0.0351$ & PO4 & $0.0218$ & $0.0211$ & $0.0256$ & $0.0228$ \\
Oz & $0.0463$ & $0.0299$ & $0.0285$ & $0.0349$ & FC2 & $0.0209$ & $0.0182$ & $0.0166$ & $0.0186$ \\
F8 & $0.0349$ & $0.034$ & $0.0324$ & $0.0338$ & CP1 & $0.0159$ & $0.02$ & $0.0188$ & $0.0182$ \\
AF3 & $0.0281$ & $0.0303$ & $0.0377$ & $0.032$ & P4 & $0.0154$ & $0.0157$ & $0.0178$ & $0.0163$ \\
Fp2 & $0.0263$ & $0.0417$ & $0.0262$ & $0.0314$ & Cz & $0.0145$ & $0.0184$ & $0.0153$ & $0.0161$ \\
AF4 & $0.0305$ & $0.0255$ & $0.0342$ & $0.0301$ & CP2 & $0.0131$ & $0.0135$ & $0.0183$ & $0.015$ \\
F3 & $0.0299$ & $0.0292$ & $0.0286$ & $0.0292$ & Fz & $0.0125$ & $0.0134$ & $0.0129$ & $0.0129$ \\
\end{tabular}
\label{tab:coefgamma}
\end{center}
\end{table}

    Next important channels are FC5 ($0.0434$), O2 ($0.0432$), P7 ($0.0388$) and FC6 ($0.0387$). Which are from emotion component regions such as the middle left and right hemispheres, as well as frontal and parietal lobes~\citep{Wang2019}. Other authors have also used these channels, such as F7, F8, and T7-FC2~\citep{Javidan2021}; FP1, T7 and T8 on $\gamma$, FC6 on $\beta$~\citep{Guo2022}; O2, T8, FC5, and P7~\citep{Dura2021}; FP1-F7~\citep{Taran2019}; F7, FC5, FC6, O2, and P7~\citep{Wang2019}.

\begin{figure}[!ht]
\begin{center}
    \centering
    \includegraphics[trim = 0 250 0 200, clip, width=0.9\linewidth]{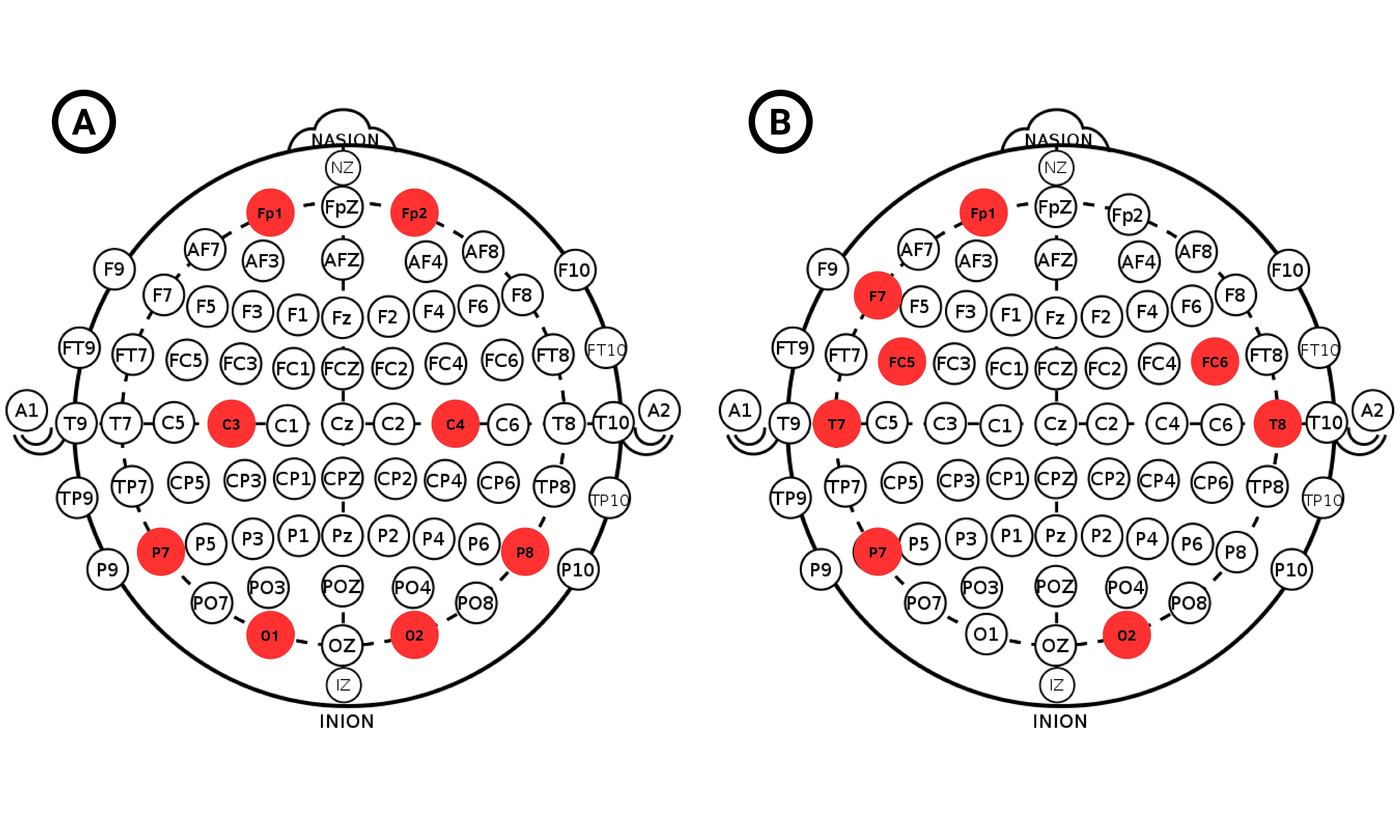}
    \caption{OpenBCI channel configurations, A) Default configuration (FP1, FP2, C3, C4, P7, P8, O1, O2), B) Optimal channel configuration for emotion recognition (Fp1, F7, FC5, FC6, T7, T8, P7, O2).}
    \label{fig:layout_final}
\end{center}
\end{figure}

    In this sense, the proposed set of 8 channels consists of 4 frontal channels (Fp1, F7, FC5, FC6), 2 temporal channels (T7, T8), 1 parietal channel (P7) and 1 occipital channel (O2); which correspond to 5 channels from the left hemisphere and 3 channels from the right hemisphere, and not including any channels from the previously established central and centro-parietal lobes. These results suggest that this set of 8 EEG channels would allow us to obtain an optimized version of the model in future iterations. Since the OpenBCI system allows channel reconfiguration, it could be easily implemented to measure EEG signals from such electrodes. So instead of the default OpenBCI channel configuration shown in Figure~\ref{fig:layout_final} A), the proposed set of channels would be used as in Figure~\ref{fig:layout_final} B).

\subsection{Model Selection}

After the channel selection analysis described in Section~\ref{sec:chansel} was performed, 42 different classifier models were trained using these 8 channels and their respective PSDs, although results from the best 15 models are shown in Table~\ref{tab:modelcomp}. The Extra-Trees model achieves the best accuracy, BA, and F1-score, as it performed significantly better than the rest of the models (95\% accuracy) on all VAD components, with only the Random Forest following behind. In order to avoid doing an expensive computation, these models were trained using an extract of the training data, taking into account every 16 steps; this allowed for model selection in a short period of time.

\begin{table}[ht]
\caption{VAD classification models evaluation comparison on Accuracy, BA, and F1-score using the best features obtained from the channel selection process.}
   \scriptsize
       \begin{tabular}{cccc}
  \hline
  &Valence&&\\ 
  \hline
  Model&Accuracy&BA&F1-Score \\
  \hline
  Extra Trees         & 0.96 & 0.95 & 0.96 \\ 
  Random Forest       & 0.92 & 0.91 & 0.92 \\    
  Bagging Class      & 0.85 & 0,84 & 0.85 \\ 
  Decision Tree       & 0.77 & 0.76 & 0.77 \\ 
  Extra Tree & 0.76 & 0.75 & 0.76 \\ 
  K-Neighbors         & 0.73 & 0.72 & 0.73 \\    
  XGBoost    & 0.72 & 0.69 & 0.71 \\ 
  NuSVC      & 0.70 & 0.67 & 0.70 \\ 
  LGBM       & 0.67 & 0.63 & 0.66 \\ 
  SVC        & 0.58 & 0.53 & 0.56 \\    
        
       \end{tabular}
       \begin{tabular}{cccc}
  \hline
  &Arousal&&\\ 
  \hline
  Model&Accuracy&BA&F1-Score \\
  \hline
  Extra Trees         & 0.96 & 0.95 & 0.96 \\  
  Random Forest       & 0.90 & 0.90 & 0.90 \\ 
  Bagging Class.      & 0.84 & 0.83 & 0.84 \\  
  Decision Tree       & 0.76 & 0.75 & 0.76 \\  
  Extra Tree & 0.74 & 0.74 & 0.74 \\  
  K-Neighbors         & 0.72 & 0.72 & 0.72 \\    
  NuSVC      & 0.72 & 0.70 & 0.72 \\ 
  XGBoost    & 0.70 & 0.68 & 0.70 \\ 
  LGBM       & 0.66 & 0.63 & 0.65 \\ 
  SVC        & 0.58 & 0.54 & 0.57 \\    
  
       \end{tabular}
       \begin{tabular}{cccc}
  \hline
  &Dominance&&\\ 
  \hline
  Model&Accuracy&BA&F1-Score \\
  \hline
  Extra Trees         & 0.96 & 0.96 & 0.96 \\  
  Random Forest       & 0.92 & 0.91 & 0.92 \\ 
  Bagging Class.      & 0.87 & 0.86 & 0.87 \\  
  Decision Tree       & 0.79 & 0.78 & 0.79 \\ 
  Extra Tree & 0.77 & 0.75 & 0.77 \\ 
  K-Neighbors         & 0.74 & 0.73 & 0.74 \\
  XGBoost    & 0.73 & 0.70 & 0.73 \\
  LGBM       & 0.70 & 0.65 & 0.69 \\
  NuSVC      & 0.68 & 0.64 & 0.68 \\  
  SVC        & 0.62 & 0.57 & 0.61 \\   
  
       \end{tabular}
       \label{tab:modelcomp}
\end{table}

In constructing the Extra-Trees model, decision trees are generated from the entire dataset. Unlike traditional tree algorithms where the best split among all possible splits is chosen, in the Extra Trees methodology, splits are randomly selected for each candidate feature, and the best of these randomly generated splits is used. When combined with ensemble techniques where multiple trees are built and averaged, this randomness often produces a model that is less prone to overfitting. Additionally, this random selection eliminates the need for bootstrapping in sampling, meaning the whole sample is used in constructing each tree~\citep{Geron}.

\subsection{Feature Selection}
In order to determine the optimal number of features, the Extra-Trees' GI was applied on all the 8-channel's PSD and index features. Furthermore, a series of Extra-Trees models were fitted using the best 25 to 35 features. Results of the accuracy of scores on Valence, Arousal, and Dominance prediction with different numbers of features with these models are shown in Table~\ref{tab:table-featuresel}. Highlighted results show the best number of features for each label and the best overall average. The average accuracy of the best number of features for each channel was calculated; there can be a decrease in average accuracy using the best 34 features than the best 35 features, so 34 features were selected as the optimal number of features to be used in the final model. In this sense, a final Extra-Trees model was fitted using the best 34 features for each VAD component.

\begin{table}[htbp]
\centering
    \caption{\label{tab:table-featuresel} Accuracy comparison of VAD predictive models using different number of features.}
\begin{tabular}{l*{4}{c} r}
Number of features & Valence & Arousal & Dominance & Average \\
\hline
25  & 0.957 & 0.959 & 0.963 &  0.960  \\
26  & 0.958 & 0.958 & 0.963 &  0.960  \\
27  & 0.960 & 0.959 & 0.964 &  0.961  \\
28  & 0.958 & 0.957 & 0.964 &  0.960  \\
29   & 0.959 & 0.958 & 0.965 &  0.961  \\
30   & 0.959 & 0.958 & \textbf{0.967} &  0.962  \\
31  & 0.958 & 0.957 & 0.965 &  0.960  \\
32   & 0.960 & \textbf{0.960} & 0.965 &  0.962  \\
33   & \textbf{0.962} & 0.957 & 0.965 &  0.961  \\
34   & \textbf{0.962} & 0.959 & \textbf{0.967} &  \textbf{0.963}  \\
35   & 0.960 & \textbf{0.960} & 0.966 &  0.962  \\
\end{tabular}\par
\end{table}

Table~\ref{tab:finalfeatures} shows the final chosen features for each of the VAD-trained models. There are a total of 34 features for each VAD classification model. Powerbands are ordered according to their most prevalent powerband in descending order. It can be seen that in the three models (VAD), the most predominant features were from $\theta$, $\gamma$, $\beta$, and $\alpha$, as delta only had 1 feature on Valence and 1 feature on Arousal. Furthermore, $\theta$, $\beta$, and $\gamma$ had the most number of channels, with 23, 24, and 24 channels, respectively, following $\alpha$ with 17 channels. It is quite interesting that the $\gamma$ bandpower is still prevalent, as shown in the channel selection analysis on Section~\ref{sec:chansel}, and that the higher frequency features are the most related to emotions, as also suggested in the same section at Table~\ref{tab:pca_r}. The hybrid feature selection method implemented, which used Pearson's linear correlation coefficient and Extra-Trees' non-linear GI, led to better generalization across the dataset, thus gathering essential insights that lead to optimized performance on 8-channel emotion recognition while including additional index features.

\begin{figure}[!ht]
\begin{center}
    \centering
    \includegraphics[trim = 0 150 0 150, width=0.75\linewidth]{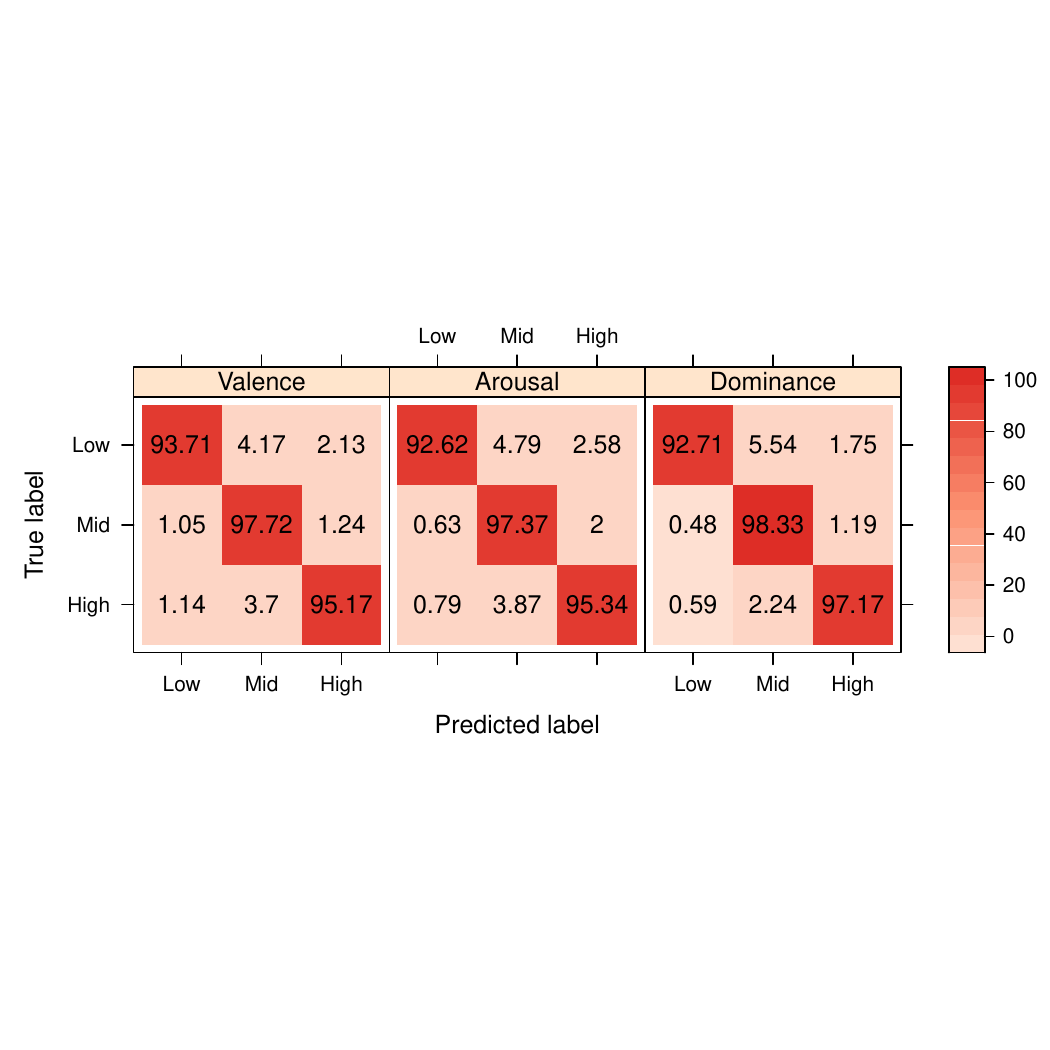}
    \caption{In confusion matrices for each VAD component model, the highest accuracy is expected to be in the diagonal element in each matrix, as it represents a correct classification of a predicted label according to the true label.}
    \label{fig:confmat}
\end{center}
\end{figure}

Best model’s performance (Extra-Trees with 34 features) achieved an accuracy of 0.96 for Valence, 0.959 for Arousal and 0.967 for Dominance, with an average accuracy of 0.963. The respective confusion  matrices are shown in Figure~\ref{fig:confmat}. It can be seen that, overall, the accuracy for each of the quadrants is $> 0.96$, and there is not a clear sign of miss-classification of each of the true and predicted labels, thus showing that residuals are randomly distributed and that high accuracy is balanced among classes of each of the models.

\begin{table}[ht]
\centering
\caption{Top 34 Features for Valence, Arousal and Dominance.}
   \footnotesize
       \begin{tabular}{cc}
Valence\\
\hline
Powerband & Channel \\
\hline
\multirow{7}{*}{Theta} & FC6 \\
                    & F7 \\
                    & Fp1 \\
                    & T8 \\
                    & FC5 \\
                     & T7 \\
                     & P7 \\
\hline
\multirow{8}{*}{Gamma} & F7 \\
                    & P7 \\
                    & Fp1 \\
                    & FC5 \\
                    & T8 \\
                    & T7 \\
                    & O2 \\
                    & FC6 \\
\hline
\multirow{7}{*}{Beta} & FC5 \\
                    & Fp1 \\
                    & FC6 \\
                    & T8 \\
                    & F7 \\
                    & T7 \\
                    & P7 \\
                    & O2 \\
\hline
\multirow{7}{*}{Alpha} & FC6 \\
                    & F7 \\
                    & T7 \\
                    & Fp1 \\
                    & FC5 \\
                    & T8 \\
                    & P7 \\
\hline
Delta & O2 \\
\hline
Engagement & FC5 \\
\hline
Fatigue & P7 \\
\hline
Excitement & P7 \\
\hline
       \end{tabular}
       \begin{tabular}{cc}
Arousal\\
\hline
Powerband & Channel \\
\hline
\multirow{9}{*}{Gamma} & O2 \\
                    & Fp1 \\
                    & T8 \\
                    & F7 \\
                    & FC5 \\
                    & T7 \\
                    & P7 \\
                    & FC6 \\
\hline
\multirow{8}{*}{Beta} & FC5 \\
                    & Fp1 \\
                    & T7 \\
                    & F7 \\
                    & P7 \\
                    & FC6 \\
                    & T8 \\
                    & O2 \\
\hline
\multirow{8}{*}{Theta} & FC6 \\
                    & Fp1 \\
                    & F7 \\
                    & T8 \\
                    & P7 \\
                    & T7 \\
                    & FC5 \\
                    & O2 \\
\hline
\multirow{4}{*}{Alpha} & P7 \\
                    & FC6 \\
                    & FC5 \\
                    & T7 \\
                    & F7 \\
                    & T8 \\
                    & Fp1 \\
                    & O2 \\
\hline
Delta & O2 \\
\hline
Engagement & O2\\
\hline
       \end{tabular}
       \begin{tabular}{cc}
Dominance\\
\hline
Powerband & Channel \\
\hline
\multirow{7}{*}{Theta} & FC6 \\
                    & F7 \\
                    & P7 \\
                    & FC5 \\
                    & T7 \\
                    & Fp1 \\
                    & T8 \\
\hline
\multirow{8}{*}{Gamma} & FC6 \\
                    & Fp1 \\
                    & T8 \\
                    & T7 \\
                    & FC5 \\
                    & P7 \\
                    & F7 \\
                    & O2 \\
\hline
\multirow{8}{*}{Beta} & FC5 \\
                    & T8 \\
                    & T7 \\
                    & Fp1 \\
                    & FC6 \\
                    & P7 \\
                    & F7 \\
                    & O2 \\
\hline
\multirow{5}{*}{Alpha} & FC6 \\
                    & FC5 \\
                    & T8 \\
                    & T7 \\
                    & Fp1 \\
\hline
\multirow{3}{*}{Engagement} & FC5 \\
                           & Fp1 \\
                           & FC6 \\
\hline
Fatigue & P7 \\
\hline
\multirow{2}{*}{Excitement} & FC5 \\
                           & FC6 \\
\hline
\end{tabular}
\label{tab:finalfeatures}
\end{table}

\section{Conclusions} \label{sec:discussion}

After all previous analyses were performed, a final model was trained. The chosen model was an Extremely Randomized Trees Classifier (Extra-Trees), as it showed significantly higher precision than the rest of the chosen models. The final features are shown in Table~\ref{tab:finalfeatures}, calculated with 5-second time windows, using the proposed channels for the optimal OpenBCI Ultracortex IV configuration at Figure~\ref{fig:layout_final} B). 

The current model is based on PSD estimations only. However, different types of features could add higher complexity, such as neural connectivity metrics. Functional connectivity metrics could be added to the model, for instance, between each channel opposite pair FC5-FC6, T7-T8, as well as other types of bipolar connections such as FP1-F7~\citep{Taran2019}, in addition to T7-P7 and T7-FP1~\citep{Meyers2021}, that have been reported previously as efficient neural markers.

It is also important to note that the model can be expanded to include variables other than EEG. Integrating other physiological measures, such as Electro-oculography (EOG), Electromyography (EMG), and Electrocardiography (ECG), can bolster the robustness and accuracy of our emotion prediction. These complementary measures provide a multifaceted view of the human emotional response, ensuring a more comprehensive analysis~\citep{Koelstra}. A future iteration of this algorithm will include such variables,  into account,  aiming to increase the complexity and accuracy of the model. 

This project presented several limitations. Firstly, the OpenBCI Ultracortex Mark IV that was employed for the real-time application only has 8 available channels for data collection. This can potentially limit the granularity and spatial resolution of the EEG data. In a future scenario, we could add a daisy extension to have more available channels. Secondly, the utilized dataset is derived from a relatively small sample size of just 32 subjects, which can introduce biases and might not be able to correctly represent the broader population, affecting the model’s overall generalizability.

Future steps of this work will also include the implementation of the real-time algorithm into experimental tests of the users while interacting with the immersive platform, and the generated emotions will be compared to the ones experienced by users in a (non-immersive) control group. Moreover, the inclusion of more biometric signals will offer more insights into the interactions experienced by test subjects when presented with these different scenarios.

\bibliographystyle{apalike}
\bibliography{references}  

\end{document}